\documentclass{llncs}

\usepackage{amsmath,gastex}
\usepackage{amssymb}
\usepackage{multirow}
\usepackage{epic,eepic,latexsym}
\usepackage[english]{babel}

\usepackage{mathrsfs}
\usepackage[latin1]{inputenc}

\newcommand{\ie}{{\it i.e.}}

\newcommand{\hf}{\hfill}

\newcommand{\debitem}{\begin{itemize} }
\newcommand{\debeqn}{\begin{eqnarray}}   
\newcommand{\debeqno}{\begin{eqnarray*}} 

\newcommand{\finitem}{\end{itemize} }
\newcommand{\fineqn}{\end{eqnarray}}   
\newcommand{\fineqno}{\end{eqnarray*}} 

\newcommand{\imp}{ \Rightarrow}
\newcommand{\equ}{ \Leftrightarrow}
\newcommand{\ti}{\times}
\newcommand{\vide}{\emptyset}
\newcommand{\inc}{\subseteq}
\newcommand{\lc}{\left<}
\newcommand{\rc}{\right>}

\newcommand{\p}{\psi}
\renewcommand{\a}{\alpha}
\renewcommand{\b}{\beta}

\newcommand{\w}{\omega}

\newcommand{\ar}{ \mathcal{A}} 
\newcommand{\br}{ \mathcal{B}} 
\newcommand{\ccc}{ \mathcal{C}} 
\newcommand{\dr}{ \mathcal{D}} 
\newcommand{\fr}{ \mathcal{F}} 
\newcommand{\gr}{ \mathcal{G}} 
\newcommand{\rr}{ \mathcal{R}} 
\newcommand{\sr}{ \mathcal{S}} 

\newcommand{\bigO}{ \mathcal{O}}

\newcommand{\defplus}[3]{+_{#2}(#1,#3)}

\newcommand{\aordinal}{\alpha}
\newcommand{\aordinalbis}{\beta}
\newcommand{\aordinalter}{\gamma}

\newcommand{\logic}[1]{{\rm LTL}(#1)}
\newcommand{\next}[1]{{\tt X}^{#1}}
\newcommand{\until}[1]{{\tt U}^{#1}}
\newcommand{\sometimes}[1]{{\tt F}^{#1}}
\newcommand{\always}[1]{{\tt G}^{#1}}

\newcommand{\set}[1]{\{ #1 \}}

\newcommand{\pair}[2]{\langle #1,#2 \rangle}
\newcommand{\triple}[3]{\langle #1,#2,#3 \rangle}

\newcommand{\powerset}[1]{2^{#1}}

\newcommand{\avarprop}{p}
\newcommand{\avarpropbis}{q} 
 
\newcommand{\varprop}{{\rm AP}} 
\newcommand{\aformula}{\phi} 
\newcommand{\aformulabis}{\psi} 

\newcommand{\amodel}{\sigma}

\newcommand{\aletter}{a}
\newcommand{\aletterbis}{b}
\newcommand{\aalphabet}{\Sigma}
\newcommand{\lang}[1]{{\rm L}(#1)}

\newcommand{\aautomaton}{{\mathcal A}}

\newcommand{\egdef}{\stackrel{\mbox{\begin{tiny}def\end{tiny}}}{=}} 

\newcommand{\step}[1]{\xrightarrow{\!\!#1\!\!}}

\newcommand {\twoexptime} {\textsc{2exptime}}

\newcommand{\todo}[1]{} 
\newcommand{\supr}[1]{}

\begin{document}
\title{Controller synthesis \& Ordinal Automata\thanks{The author 
acknowledges partial support by the ACI ``S{\'e}curit{\'e} et Informatique'' 
CORTOS. \url{http://www.lsv.ens-cachan.fr/aci-cortos/}  }}
\author{Thierry Cachat}
\institute{LIAFA/CNRS UMR 7089 \& Universit{\'e} Paris 7, France
}

\maketitle

\begin{abstract} 
Ordinal automata are used to model physical systems with Zeno behavior.
Using automata and games techniques we solve a control problem formulated
and left open by Demri and Nowak in 2005. It involves partial observability
and a new synchronization between the controller and the environment.
\end{abstract}

%
%
\section{Introduction}
%

\paragraph{Controller synthesis.}
The synthesis of controller is today one of the most important challenges
in computer science. Since \cite{RamadgeWonham89} different formalisms have
been considered to model (un)controllable and (un)observable actions. 
The problem is well understood for finite systems 
admitting infinite behavior (indexed by $\w$) \cite{Pnueli&Rosner89}.
Recent developments concern extensions to e.g. infinite state systems 
or timed systems \cite{BDMP03}. 

Transforming control problems into two-player games have provided
efficient solutions \cite{Thomas95}. 
In this setting the controller is modeled by a player
and the environment by her opponent. Determining whether a controller exists 
falls down to determine the winner and computing a winning strategy is equivalent
to synthesizing a controller.

\paragraph{Ordinal automata.}
A B{\"u}chi or Muller automaton, after reading an $\w$-sequence, simply accepts or 
rejects, depending on the states visited infinitely often. In an ordinal
automaton there is a limit transition to a new state, also depending on the
states visited infinitely often and the run goes on from this state. This
allows to model a system preforming $\w$ actions in a finite time and reaching
a limit state.

\paragraph{Systems with Zeno behaviors.}
When modeling physical systems we face the problem that different components can
have different time scales. For example the controller of an anti-lock braking 
system (ABS) is supposed to react much quicker than the physical environment.
In the opposite one can consider physical systems admitting Zeno behavior
---infinitely many actions in a finite amount of time--- whereas the controller 
is a computer
with constant clock frequency. A simple example is a bouncing ball. 
Another one is the physical
description of an electronic circuit which evolves much quicker than its logical 
description in VHDL. The speeds are so different that one can consider that the
former one evolves infinitely quicker than the latter one.

Following this idea Demri and Nowak \cite{DemriNowak05} have 
proposed to model physical
systems by ordinal automata, thus admitting ordinal sequences as behavior
(typically of length $\w^k$). They
define a logic $\logic{\w^k}$ as an extension of ${\rm LTL}$ to express properties
of such systems. The controller should be a usual automaton whose execution is an
$\w$-sequence. The synchronization between controller and environment is the
following: environment makes $\w^{k-1}$ steps ``alone'', then controller and
environment makes one step together, and so on.

Particularly in the context of timed systems, different techniques have been
proposed to forbid or restrict Zeno behaviors, see introduction of 
\cite{AFH+03} for an overview. Our claim is that we want to allow Zeno behavior,
to model them and express properties about them, and finally to control such systems.

\paragraph{Our contribution.} 
The main contribution of our article is a solution to the control problem stated and 
left open in \cite{DemriNowak05}. Given a physical system modeled by an ordinal 
automaton and a formula $\p$ of $\logic{\w^k}$ we want to determine whether a 
controller exists and synthesize one. The technique used is to transform the control
problem into a game problem. Because of the unobservable actions and also because of
the different time scales, the controller can not fully observe the current state of
the system. For that reason we construct a game of imperfect information. Another
difficulty is that the length of the interaction is greater than $\w$, but 
fortunately one can
summarize $\w^{k-1}$ steps done by the environment ``alone''.
Several games and automata techniques are used.

\paragraph{Related work.}
It is known that games of imperfect information have higher computational 
complexity \cite{Reif84}. Zeno behavior have already been considered in the
literature. In \cite{BerardPicaronny} languages of ordinal words accepted by
timed automata are studied. 
In the framework of hybrid systems 
\cite{AsarinMaler,MR1650864} 
or cellular automata on continuous time and space \cite{Durand-Lose} 
it is known that allowing Zeno behaviors gives rise to highly
undecidable problems. In \cite{DemriNowak05} Demri and Nowak solve 
the satisfiability and the model-checking problem for $\logic{\w^k}$: 
given an ordinal automaton reading $\w^k$-sequences and a 
formula $\p$, determine whether every run of the automaton 
satisfies $\p$. For this they use a ``succinct'' form of ordinal automata to have
better complexity bounds.

\paragraph{Plan of the paper}
In the next section we present the temporal logic $\logic{\w^k}$, ordinal automata
and the control problem. 
We show a translation to first order logic.
In Section~\ref{sec:3} we solve our main problem. We first explain 
how to translate
it to a game and why the controller has imperfect information about the system.
An example is provided in Section~\ref{sec:4}.

%
%
\section{Reasoning about transfinite sequences}
%

We assume basic knowledge about ordinals less than $\omega^\omega$,
see e.g.~\cite{Rosenstein82}.
An {\em ordinal} is a well and totally ordered set.
It is either $0$ or a successor ordinal of the form $\beta+1$ or a limit ordinal.
The first limit ordinal is denoted $\w$. For all ordinal $\a$,
$\b<\a \equ \b\in \a$ and $\a=\{\b : \b<a\}$.
In this article we restrict ourselves to ordinals less or equal than 
$\w^\w$. By the Cantor Normal Form theorem, for all $\a<\w^\w$ there exists
unique integers $p,n_1,\dots,n_p$ and $k_1,\dots,k_p$ such that 
$k_1 >k_2 >\dots >k_p$ and 
$\a=\w^{k_1}n_1+\w^{k_2}n_2+\dots+  \w^{k_p}n_p$. Recall e.g. that
$2\w=\w$ and $\w+\w^2=\w^2$.
An ordinal $\aordinal$ is said to be closed under addition whenever $\aordinalbis,
\aordinalbis' < \aordinal$ implies $\aordinalbis + \aordinalbis' <
\aordinal$. 
In particular for every $\aordinal \leq \omega^{\omega}$, 
$\aordinal$ is closed under addition
iff $\aordinal$ is equal to $\omega^{\aordinalbis}$ for some $\aordinalbis \leq \omega$
or $\aordinal=0$.
%
In the following we will consider a logic whose models are $\w^k$ sequences for
some $k<\w$.

\subsection{Temporal Logic}
%

We recall the definition of the logic $\logic{\a}$ introduced in 
\cite{DemriNowak05}.
For every ordinal $\aordinal$ closed under addition, the models of
$\logic{\aordinal}$ are precisely sequences of the form
$\amodel: \aordinal \rightarrow \powerset{\varprop}$ for some countably infinite set
$\varprop$ of atomic propositions.
The formulas of $\logic{\aordinal}$ are defined as follows:
$
\aformula ::= \avarprop \ \mid \ 
              \neg \aformula \ \mid \ 
              \aformula_1 \wedge \aformula_2 \ \mid \
              \next{\aordinalbis} \aformula \ \mid \ 
              \aformula_1 \until{\aordinalbis'} \aformula_2
$, 
where $\avarprop \in \varprop$, $\aordinalbis < \aordinal$ and $\aordinalbis' \leq \aordinal$.
 The satisfaction relation is inductively defined below where $\amodel$ is a model
for $\logic{\aordinal}$ and $\aordinalbis < \aordinal$:
\begin{itemize}
\itemsep 0 cm
\item $\amodel, \aordinalbis \models \avarprop$ iff $\avarprop \in \amodel(\aordinalbis)$,
\item  $\amodel, \aordinalbis \models \aformula_1 \wedge \aformula_2$ iff 
        $\amodel, \aordinalbis \models \aformula_1$ and $\amodel, \aordinalbis \models \aformula_2$, \ 
       $\amodel, \aordinalbis \models \neg \aformula$ iff not $\amodel, \aordinalbis \models \aformula$,
\item $\amodel, \aordinalbis \models \next{\aordinalbis'} \aformula$ iff
       $\amodel, \aordinalbis +  \aordinalbis' \models \aformula$,
\item $\amodel, \aordinalbis \models \aformula_1 \until{\aordinalbis'} \aformula_2$ iff
      there is $\aordinalter < \aordinalbis'$ such that 
       $\amodel, \aordinalbis +  \aordinalter \models \aformula_2$ and
      for every $ \aordinalter' <  \aordinalter$, $\amodel, \aordinalbis +  \aordinalter' \models \aformula_1$.
\end{itemize}
Closure under addition of $\aordinal$ guarantees that $\aordinalbis +  \aordinalbis'$ 
and $\aordinalbis +  \aordinalter$ above are strictly smaller than $\aordinal$.
Usual ${\rm LTL}$ is expressively equivalent to $\logic{\w}$: 
$\next{}$ is equivalent to $\next{1}$ and $\until{}$ is equivalent to $\until{\w}$,
conversely $\next{n}$ and $\until{n}$ can be expressed in ${\rm LTL}$.
Standard abbreviations are also extended: $\sometimes{\aordinalbis} \aformula \egdef
\top \until{\aordinalbis} \aformula$ and $\always{\aordinalbis} \aformula \egdef \neg \sometimes{\aordinalbis} \neg \aformula$.
Using Cantor Normal Form it is easy to effectively encode an $\logic{\w^k}$ formula
for $k<\w$. 
We provide below properties  dealing with limit states that can be 
easily expressed in $\logic{\omega^k}$ ($k \geq 2$). \\
``$\avarprop$ holds in the states indexed by limit ordinals strictly less than $\omega^k$'':
      $$\always{\omega^k}(\next{\omega} \avarprop \wedge \cdots \wedge \next{\omega^{k-1}} \avarprop).$$
For $1 \leq k' \leq k-2$, ``if $\avarprop$ holds infinitely often in states indexed by ordinals of the
      form $\omega^{k'} \times n$, $n \geq 1$, then $\avarpropbis$ holds in the state indexed by $\omega^{k'+1}$'':
      $$ (\always{\omega^{k'+1}}  \sometimes{\omega^{k'+1}} \next{\omega^{k'}} \avarprop) \Rightarrow (\next{\omega^{k'+1}} q).$$

\subsection{Translation to First Order Logic}
%

In \cite{DemriNowak05} it is proved that $\logic{\w^\w}$ 
(hence also $\logic{\w^k}$) can be translated to the
monadic second order theory of $\lc \w^\w,<\rc$, which gives a 
non-elementary decision procedure
for satisfiability \cite{Buechi73}. We improve this result by showing that
$\logic{\w^\w}$ can be translated even to the first order theory (FO) of 
$\lc \w^\w,<\rc$.

\begin{proposition}
For every $\logic{\w^\w}$ formula there exists an equivalent first order formula 
over $\lc \w^\w,<\rc$.
\end{proposition}
It is open whether the converse also holds, extending Kamp's theorem \cite{Kamp68}.

\begin{proof}[sketch] 
The main point is the definition of a formula $\defplus{x}{\aordinalbis}{y}$ for some
$\aordinalbis < \omega^{\omega}$ such that 
$\pair{\omega^{\omega}}{<} \models_{v} \defplus{x}{\aordinalbis}{y}$ with $v:\set{x,y} \rightarrow
\omega^{\omega}$ iff $v(y) = v(x) + \aordinalbis$.
The relation $\models_{v}$ is the standard satisfaction relation under the valuation $v$.
The formulas of the form $\defplus{x}{\aordinalbis}{y}$ with $\aordinalbis < \omega^{\omega}$ are inductively defined as: 
\begin{enumerate}
\itemsep 0 cm
\item $\defplus{x}{0}{y} \egdef (x = y)$ , 
\item $\defplus{x}{1}{y} \egdef (x < y) \wedge \forall \ z \ (z > x \Rightarrow y \leq z)$ ,
\item $\defplus{x}{\omega^k n + \aordinalbis}{y} \egdef
       \exists \ z \ \defplus{x}{\omega^{k}}{z} \wedge \defplus{z}{\omega^k (n-1) + \aordinalbis}{y}$
          ($n \geq 1$, $k \geq 0$) ,
\item 
	 $+_{\w^k} (x,y) \egdef
	(x<y)\wedge\forall z(x\leq z< y \imp \exists z'(+_{\w^{k-1}}(z,z')\wedge z'<y))\wedge\\
\forall y'[((x<y')\wedge\forall z(x\leq z< y'\imp\exists z'(+_{\w^{k-1}}(z,z')\wedge z'<y'))) \imp y\leq y']$ ($k \geq 1$) .
\end{enumerate}
For $k=1$, the latter formula is written in the following way.
The ordinal $y$ such that $+_{\w} (x,y)$ holds is greater than $x$, 
greater than every finite step successors of $x$, and $y$ is the least ordinal
satisfying this two conditions. By induction one can show that $y>x+n$ for
every $n<\w$.
Analogously for $k>1$, the formula implies that $y>x+\w^{k-1} n$ for
every $n<\w$.
\qed
\end{proof}

The first order theory of $\lc \w^\w,+\rc$ has a non-elementary
decision procedure \cite{Maurin96}. We are not aware of the exact complexity
of the more restricted first order theory of $\lc \w^\w,<\rc$.
We use ordinal automata, both to model
physical systems and to represent specifications.

\subsection{Ordinal Automata}
%

Since Büchi in the 1960s and Choueka in the 1970s, different forms of 
ordinal automata have been proposed.
A particular class of ordinal automata is well suited to solve our problem.
See \cite{Bedon98} for the equivalence between different definitions.
Ordinal automata has two kinds of transitions: usual one-step transition for
successor ordinals and limit transitions for limit ordinals where the state
reached is determined by the set of states visited again and again ``before''
that ordinal.
An ordinal automaton is  a tuple $(Q, \aalphabet, \delta, E, I, F)$ where:
\begin{itemize}
\item $Q$ is a finite set of states,
\item $\Sigma$ is a finite alphabet,
\item $\delta \; \subseteq \; Q \times \aalphabet \times Q$
is a one-step transition relation,
\item $E \; \subseteq \; \powerset{Q} \times Q$ is a limit transition relation,
\item $I \subseteq Q$ is a finite set of initial states,
\item $F \subseteq Q$ is a finite set of final states.
\end{itemize}

We write $q \step{\aletter} q'$ whenever $\triple{q}{\aletter}{q'} \in \delta$
and $P\step{}q$ whenever $\pair{P}{q}\in E$.
%
A {\em path} of length $\aordinal + 1$ is an $(\aordinal + 1)$-sequence 
$r: \aordinal + 1 \rightarrow Q$ labeled by an $\aordinal$-sequence 
$\amodel: \aordinal \rightarrow \aalphabet$ such that for every 
$\aordinalbis \in \aordinal$, 
$r(\aordinalbis) \step{\amodel(\aordinalbis)} r(\aordinalbis+1)$ and
for every limit  ordinal $\aordinalbis \in \aordinal +1$, there is 
      $P \step{} r(\aordinalbis) \in E$ s.t.
      $P = \mbox{\it cofinal}(\aordinalbis,r)$  with
      $
      \mbox{\it cofinal}(\aordinalbis,r) \egdef
      \{q \in Q: {\rm for \ every} \ \aordinalter \in \aordinalbis, \ {\rm there \ is} \
           \aordinalter' \ {\rm such \ that} 
        \ \aordinalter < \aordinalter' < \aordinalbis \ {\rm and} \
           r(\aordinalter') = q\}$.
The set $\mbox{\it cofinal}(\aordinalbis,r)$ is the set of states visited again and
again arbitrary close to $\aordinalbis$ (hence infinitely often).
\\
If moreover $r(0) \in I$, it is a {\em run}. If moreover $r(\aordinal) \in F$,
it is accepting.
\\[1 em]
\begin{minipage}[b]{9 cm}
\begin{example}  
We present here an example of ordinal automaton $\aautomaton$
with limit transitions $\set{0} \step{} 1$ and $\set{0,1} \step{} 2$.
One can show that $\lang{\aautomaton}$ contains only $\omega^2$-sequences
and $\lang{\aautomaton} = (\aletter^{\omega} \cdot \aletterbis)^{\omega}$.
\end{example}
\end{minipage}
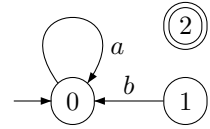
\begin{picture}(30,16)(-2,2)
\gasset{Nadjust=wh,Nadjustdist=2}
\node[Nmarks=i](s0)(10,5){$0$}
\node(s1)(25,5){$1$}
\node[Nmarks=r](s2)(25,15){$2$}
\drawedge[ELside=r](s1,s0){$\aletterbis$}
\drawloop[loopangle=90,ELpos=80](s0){$\aletter$}
\end{picture}
\\[1 em]
For all $k<\w$ there exists an ordinal automaton accepting exactly the sequences of
length $\w^k$, using $k+1$ states. But if an ordinal automaton accepts a sequence
of length $\w^\w$, then it must also accept longer sequences. That is a second 
reason, beside closure under addition, why we restrict ourselves to ordinals less 
than $\w^\w$.

\paragraph{Level}
An ordinal automaton $\aautomaton = \triple{Q,\aalphabet}{\delta,E}{I,F}$
is of {\em level} $k \geq 1$ iff there is a map $l: Q \rightarrow
\set{0, \ldots,k}$ such that:
\begin{itemize}
\itemsep 0 cm
\item for every $q \in F$, $l(q) = k$;
\item $q \step{a} q' \in \delta$ implies $l(q') = 0$ and $l(q) < k$;
\item $P \step{} q \in E$ implies
$l(q) \geq 1$, for every $q' \in P$, $l(q') < l(q)$, and
there is $q' \in P$ such that $l(q') = l(q) -1$.
\end{itemize} 

The idea is that a state of level $i$ is reached at positions $\b+\w^i.j$, $j<\w$.
Since \cite{VardiWolper86}, different techniques for translating logic formulas 
to automata are widely used.

\begin{proposition}[\cite{DemriNowak05}] For all $\logic{\w^k}$ formula, there
exists an equivalent ordinal automaton.
\end{proposition}
This result can be obtain by translating an $\logic{\w^k}$ formula into an
equivalent first order formula (or even monadic second order) and applying 
results from \cite{Buechi73}. In \cite{DemriNowak05} a succinct version of 
ordinal automata is defined to improve the complexity of the translation from 
non-elementary to polynomial (resp. exponential) space when integers in the 
formulas are encoded in unary (resp. binary).

\subsection{Control Problem}\label{sec-CProblem}
%

Before we recall the control problem from \cite{DemriNowak05}
we need some preliminary definitions. In order for the physical system to 
evolve much faster than the controller we need a particular synchronization
between them.

\paragraph{Synchronous product.}
We define below the synchronous product of two ordinal automata having possibly
different alphabets. They synchronize only on the common actions.
This is used later to model unobservable actions.
Let $\Sigma_i=2^{Act_i}$ for $i=1,2$, a letter from $\Sigma_i$ is a set of actions.
Given two ordinal automata
$\aautomaton_i = \triple{Q_i, \Sigma_i}{\delta_i, E_i}{I_i, F_i}$, for $i=1,2$,
their synchronous product is defined as 
$\aautomaton_1 \times \aautomaton_2 =
\triple{Q, \Sigma}{\delta, E}{I, F}$ where:
\begin{itemize}
\itemsep 0 cm
\item
$Q = Q_1 \times Q_2$,\qquad  $\Sigma = 2^{Act_1\cup Act_2}$.
\item
$\pair{q_1}{q_2} \overset{a}{\longrightarrow} \pair{q'_1}{q'_2} \in \delta$\ \ iff\ \
$q_1\step{a\cap Act_1} q'_1$ and $q_2\step{a\cap Act_2} q'_2$.
\item
$P \step{} \pair{q_1}{q_2} \in E$ iff
there exists $P_1 \step{} q_1 \in E_1$ and
$P_2 \step{} q_2 \in E_2$ such that
$\{ q : \pair{q}{q'} \in P \} = P_1$ and
$\{ q': \pair{q}{q'} \in P \} = P_2$.

\item
$I = I_1 \times I_2$,\qquad  $F = F_1 \times F_2$.
\end{itemize}

\paragraph{Lifting.}
In order to synchronize the system with a controller working on $\omega$-sequences,
we need to transform the controller so that 
its product with $\mathcal{S}$ only constraints states
 on positions $\omega^{k-1} \times n$, $n < \w$.
The other positions are not constrained.

Let $\aautomaton = \lc Q, \Sigma,\delta, E,I, F,l\rc$ be an automaton of level 1.
We define its lifting
$\mbox{\it lift}_k(\aautomaton)$
at level $k \geq 2$ to be the automaton
$\lc Q', \Sigma,\delta', E',I', F',l'\rc$ by:
\begin{itemize}
\itemsep 0 cm
\item
$Q' = \{0, \ldots, k \} \times Q$,\qquad
$I' = \set{k-1} \times I$,\qquad $F'=\{k\}\ti F$
\item $l'(\pair{i}{q'}) = i$,
\item
$
\delta' =
\begin{array}[t]{l}
\{
\pair{k-1}{q} \step{a}  \pair{0}{q'} \; : \;
q \overset{a}{\longrightarrow} q' \in \delta
\}
\cup
\\
\{
\pair{i}{q} \step{a} \pair{0}{q}  \; : \;
0 \leq i \leq k-2, \; a \in \Sigma, \; q \not\in F
\},
\end{array}
$
\item
$ E' =
\{
\{ \pair{0}{q}, \ldots, \pair{i-1}{q} \} \step{} \pair{i}{q} \; : \;
   1 \leq i < k, \; q \in Q
\}
\cup
\{
\{ \pair{0}{q_1}, \ldots, \pair{k-1}{q_1}, \ldots, \pair{0}{q_n}, \ldots, \pair{k-1}{q_n} \}
\step{} \pair{k}{q} \; \mid \;
\{ q_1, \ldots q_n \} \step{} q \in E
\}
$.
\end{itemize}

\begin{example}
We present below an example of ordinal automaton $\aautomaton$
with limit transition $\set{q_0, q_1} \step{} q_2$ 
and the corresponding automaton 
$\mbox{\it lift}_2(\aautomaton)$ with limit transitions
$\set{\pair{0}{q_0}} \step{} \pair{1}{q_0}$,
$\set{\pair{0}{q_1}} \step{} \pair{1}{q_1}$,
and \\
 $\set{\pair{0}{q_0}, \pair{1}{q_0}, 
       \pair{0}{q_1}, \pair{1}{q_1},} \step{} \pair{2}{q_2}$.
We omit useless transitions.
\begin{center}
\begin{picture}(120,30)(0,3)
\node[Nframe=n](A)(17,30){$\aautomaton$}
\node[Nframe=n](lift)(82,30){$\mbox{\it lift}_2(\aautomaton)$}
\gasset{Nadjust=wh,Nadjustdist=2,curvedepth=1}
\node[Nmarks=i](s0)(10,20){$q_0$}
\node(s1)(25,20){$q_1$}
\node[Nmarks=r](s2)(17,8){$q_2$}
\drawedge(s0,s1){$\aletter$}
\drawedge(s1,s0){$\aletterbis$}

\gasset{curvedepth=0}
\node[Nmarks=i](t10)(60,20){$\pair{1}{q_0}$}
\node(t01)(82,20){$\pair{0}{q_1}$}
\node(t11)(60,8){$\pair{1}{q_1}$}
\node(t00)(82,8){$\pair{0}{q_0}$}
\drawedge(t10,t01){$\aletter$}
\drawedge(t11,t00){$\aletterbis$}
\drawloop[loopangle=0](t01){$\Sigma$}
\drawloop[loopangle=0](t00){$\Sigma$}

\node[Nmarks=r](s2)(113,14){$\pair{2}{q_2}$}
\end{picture}
\end{center}
\end{example}

\begin{proposition}[\cite{DemriNowak05}]
For all $w \in \Sigma^{\omega^k}$,
$w \in \lang{\mbox{\it lift}_k(\aautomaton)}$ iff
the word $w' \in \Sigma^\omega$,
defined by $w'(i) = w(\omega^{k-1} \times i)$,
is in $\lang{\mathcal A}$.
\end{proposition}
A physical system $\mathcal{S}$ is modeled as a structure
$$\triple{\aautomaton_\sr}{\mbox{\it Act}_{\mbox{\it c}}}{\mbox{\it Act}_{\mbox{\it o}},
\mbox{\it Act}}$$
where  
$\aautomaton_\sr$ is an ordinal automaton of level $k$ with alphabet
      $\powerset{\mbox{\it Act}}$ where $\mbox{\it Act}$ is a finite set of actions,
$\mbox{\it Act}_{\mbox{\it o}} \subseteq \mbox{\it Act}$ is the set of
	observable actions,
$\mbox{\it Act}_{\mbox{\it c}} \subseteq \mbox{\it Act}_{\mbox{\it o}}$ is
      the set of controllable actions. The set 
	$\mbox{\it Act}\backslash \mbox{\it Act}_{\mbox{\it c}}$
	of uncontrollable actions is denoted
      by $\mbox{\it Act}_{\mbox{\it nc}}$.
A specification of the system $\mathcal{S}$ is naturally an $\logic{\omega^k}$ formula $\aformulabis$.
A controller $\mathcal{C}$ 
for the pair $\pair{\mathcal{S}}{\aformulabis}$ is a system whose complete executions
are $\omega$-sequences (typically ordinal automata of level 1) verifying the properties below.
\begin{description}
\itemsep 0 cm
\item{(obs)} Only observable actions are present in the controller. Hence, thanks to the synchronization mode,
      in the product system between  $\mathcal{S}$ and $\mathcal{C}$, unobservable actions do not
      change the $\mathcal{C}$-component of the current state. So the alphabet of $\mathcal{C}$
      is $\powerset{\mbox{\it Act}_{\mbox{\it o}}}$. Moreover for every state $q$ of $\mathcal{C}$
      there is a transition $q \step{\vide} q$.
\item{(unc)} From any state of $\mathcal{C}$, uncontrollable actions can always be executed:
      $
\forall q \; \cdot \;
\forall a \subseteq
\mbox{\it Act}_{\mbox{\it o}}  \setminus \mbox{\it Act}_{\mbox{\it c}}$, there is a transition
$q \step{b} q'$ in $\mathcal{C}$ such that 
$b \cap \mbox{\it Act}_{\mbox{\it nc}} = a$.
\item{(prod)} Finally, the system $\mathcal{S}$ controlled by $\mathcal{C}$ satisfies 
      $\aformulabis$. Because $\mathcal{S}$ and $\mathcal{C}$ work on sequences of different 
      length, the controlled system is in fact equal to
      $\mbox{\it lift}_{k}({\mathcal C}) \; \times \;{\mathcal S}$. 
      So $\mbox{\it lift}_{k}({\mathcal C}) \; \times \;{\mathcal S} \models \aformulabis$ should hold.
      This is equivalent 
      to the emptiness of the language of the product automaton
$
\mbox{\it lift}_{k}({\mathcal C}) \; \times \;
{\mathcal S} \; \times \; \aautomaton_{\neg \aformulabis}
$.
\end{description} 
We say that $\mathcal{C}$ is a controller for $\mathcal{S}$ (without mentioning
$\aformulabis$) if $\mathcal{C}$ fulfills the first two conditions. The notion
of final state is not relevant for the controller or the physical system. To
conform with previous definitions we require that every $(\w+1)$-run of the
controller and $(\w^k +1)$-run of $\mathcal{S}$ end in a final state.

The {\em control problem for $\logic{\omega^k}$} is defined as follows:
\\
{\bf input:} a system $\mathcal{S}=\triple{\aautomaton_\sr}{\mbox{\it Act}_{\mbox{\it c}}}{\mbox{\it Act}_{\mbox{\it o}},\mbox{\it Act}}$ with 
ordinal automaton $\aautomaton_\sr$ of level $k$ and an $\logic{\omega^k}$ formula $\aformulabis$
over atomic formulas in $\mbox{\it Act}$.
\\
{\bf output:} an ordinal automaton $\mathcal{C}$ of level 1 satisfying the conditions 
(obs), (unc) and (prod) above if there exists one. Otherwise the answer ``no controller
exists''.

%
%
\section{Solving the Control Problem} \label{sec:3}
%

Given a physical system $\sr$ modeled by an ordinal automaton $\ar_\sr$ of
level $k$ and an LTL($\w^k$)-formula $\p$, we are looking for a controller
$\ccc$ such that $\mbox{\it lift}_k(\ccc)\ti \ar_\sr \models \p$ and $\ccc$ has the expected
properties about uncontrollable and unobservable actions. 

\paragraph{From Control Problem to Game.}
Let $\br=\mbox{\it lift}_k(\ccc)\ti \aautomaton_\sr \ti \ar_{\neg \p}$.
At a given point in a run of $\br$ the controller is in a state $q$.
From $q$ and for all $o\inc Act_{o}\cap Act_{nc}$ it must have at least one
transition labeled by $o\cup c$ for some $c\inc Act_{c}$. The most general form
of a controller (possibly with infinite memory) is a function 
$f:(2^{Act_{o}})^*\ti (2^{Act_{o}\cap Act_{nc}}) \step{} 2^{Act_{c}}$, 
because the current state
of the controller shall only depend on the past observable actions. This function 
is exactly a strategy in a game that we will define. A controller for
$\pair{\mathcal{S}}{\aformulabis}$ is such that every run according to $f$ is 
winning.

Let $\ar=\ar_\sr \times \ar_{\neg \p}$. 
It is also an ordinal automaton of level $k$ :
$\aautomaton = \triple{Q, \Sigma}{\delta, E}{I, F,l}$.
We are looking for a controller ${\mathcal C}$ such that the language of
$\mbox{\it lift}_{k}({\mathcal C}) \; \times \; \ar$ is empty.
We will consider a game where the environment tries to build an accepting run of 
$\ar$, whereas the controller tries to avoid that, using the controlled actions. 
In fact the environment plays both for the system $\sr$ and for the automaton 
of $\neg \p$, as we will see later.

\subsection{Some Definitions from Game Theory}\label{sec-jeux} 
%

We recall some definitions about games. See for example \cite{Thomas95,Graedel&Th&W02}
for an introduction. An {\em arena}, or {\em game graph}, is a triple $(V_0,V_1,G)$,
where $V=V_0\cup V_1$ is the set of vertices and $G\inc V\ti V$ is the set of edges.
The vertices of $V_0$ belongs to Player~0, those of $V_1$ to Player~1 ($V_0\cap V_1=\vide$).
A {\em play} from $v_0\in V$ proceeds as follows: if $v\in V_0$, Player~0 chooses a successor
$v_1$ of $v_0$, else Player~1 does. Again from $v_1\in V_i$, Player~$i$ chooses
a successor $v_2$ of $v_1$, and so on.

A play $\pi=v_0,v_1,v_2,\dots$ is a finite or infinite sequence of vertices such that
$\forall i, (v_i,v_{i+1})\in G$. If the play is finite, the convention is that the player
who belongs the last vertex loses (he is stuck). If the play is infinite, the winner is
determined by a {\em winning set}, $Win\inc V^\omega$: Player~0 wins an infinite play
$\pi$ if and only if $\pi\in Win$. Usually $Win$ is an $\omega$-regular set, defined by
a B{\"u}chi, Rabin, parity or Muller automaton. One speaks also of {\em winning condition}. 
A {\em game} $(V_0,V_1,G,Win)$ is an arena together with a winning condition and possibly
an initial vertex $v_0\in V$.

For a game or an automaton, a B{\"u}chi condition is given by a set $F\inc V$ of ``final''
vertices and $\pi\in Win$ if and only if $\forall i>0, \exists j>i, \pi_i\in F$. A Muller
condition is given by $\fr \inc 2^V$, $\fr=\{F_1,\cdots,F_n\}$, and $\pi\in Win$ if and
only if the set of states visited infinitely often along $\pi$ is equal to one of the
$F_i$'s.

A {\em strategy} for Player~0 is a (partial) function $f_0:V^*V_0\mapsto V$ such that for every
prefix $v_0, v_1, v_2, \cdots v_i$ of a play, where $v_i\in V_0$, $f(v_0 v_1 v_2 \cdots v_i)$
is a vertex $v_{i+1}$ such that $(v_i,v_{i+1})\in G$. A play $\pi$ is played according to
a strategy $f_0$ if $\forall i, v_i\in V_0 \imp v_{i+1}=f(v_0 v_1 v_2 \cdots v_i)$. A
strategy for Player~1 is defined analogously. A strategy of Player~0 is {\em winning} if every 
play according to it is winning for Player~0.
An important case in practice is when the strategy is {\em positional}: it depends only on the
current vertex, not on the past of the play, \ie, for all $v_0, v_1, v_2, \cdots v_i$,
$f(v_0 v_1 v_2 \cdots v_i)=f(v_i)$.

From \cite{Martin75} we know that every zero-sum two-player turn based game of complete 
information with Borel winning condition (including $\omega$-regular and many more)
is determined: from a given initial configuration, one of the players has a winning strategy.

In the case of incomplete information, the players do not in general know exactly 
the current position of the game. They only know that the position belongs to a 
certain set of uncertainty. The move chosen by a player (by his strategy) shall
depend on this set, but not on the precise position of the play. As we will see in some
cases one can transform such a game into a game of complete information, where a
vertex represents a set of positions of the original game.

\subsection{A Solution With Incomplete Information}
%

\paragraph{Summarizing $\w^{k-1}$ steps.}
From the definition of $\mbox{\it lift}_k$ we see that the controller can act only
every $\w^{k-1}$ steps of the environment. Our aim is to summarize $\w^{k-1}$ 
steps of the environment in a single step. One can compute a relation
$\rr\inc Q\ti 2^Q\ti Q$ such that $(q,P,q')\in\rr$ iff there exists in
$\ar$ a path from $q$ to $q'$ of length $\w^{k-1}+1$ where the set of states
seen along this path is exactly $P$.
Note that to determine $\rr$, one has to look for cycles in $\ar$ and states
that are seen infinitely often, but in $\rr$ itself we only need to know states 
that are ever visited. The reason is that (considering 
$\mbox{\it cofinal}(\w^k,r)$ ) it is not relevant to know that some state
is visited infinitely often between e.g. $\w^{k-1}3$ and $\w^{k-1}4$ and no
more visited after $\w^{k-1}4$.
Relation $\rr$ can be computed in time $2^{\bigO (|Q|)}$ \cite{Carton02}.

\paragraph{Game.}
We introduce a game ($\gr$) modeling the interaction between 
the controller (Cont) and the environment (Env).
It is not possible in general for Cont to know exactly the current 
state of the system for several reasons.
\debitem
\item Cont cannot know the $\w^{k-1}$ steps done by the environment without control. 
\item As Env act, by choosing $v\inc Act_{nc}$, Cont can only observe the actions that
are in $Act_{o}$.
\item Moreover $\ar$ is not necessarily deterministic. 
In particular it is possible that $\ar_{\neg \p}$ is not deterministic and
Env has to ``choose'' which subformulas of $\neg \p$ he wants to make true.
\item Also Cont cannot know exactly the initial state chosen by Env.
\finitem
In the game $\gr$ Cont has partial information:
a position of the game is a subset $Q_i$ of $Q$, such that Cont knows that the 
current state of the system is in $Q_i$, but does not know which state exactly.
The game is defined by the following steps:
\begin{enumerate}
\item $i=0$ and the initial position is $Q_0=I$, the set of initial states of $\ar$
\item Env chooses $o_i\inc Act_{o}\cap Act_{nc}$,
\item Cont chooses $c_i\inc Act_{c}$,
\item there is a one step transition to 
	$$Q'_i = \{ q'\in Q : \exists u\inc Act\backslash Act_o, 
	\exists q\in Q_i, q\step{c_i\cup o_i\cup u} q'\} ,$$
\item there is a jump to $Q_{i+1}$, summarizing $\w^{k-1}$ steps 
	$$Q_{i+1} = \{ q\in Q : \exists q'\in Q'_i, \exists (q',P,q)\in\rr\} ,$$ 
\item $i=i+1$, continue at point 2.
\end{enumerate}
In this game the knowledge of Cont about the current state is exactly what a
controller can compute in the original problem, based on the observable actions.
A play is essentially a sequence $Q_0,Q'_0,Q_1,Q'_1,\dots$ (a more precise
definition of the game graph is given below) and now it is more intricate to 
determine the winner. The sequence $Q_0,Q'_0,Q_1,Q'_1,\dots$ represents the 
point of view of the controller, and we call it an abstract play.
After the game is played a {\em referee} has 
to choose inside this abstract play a concrete path (if it exists one)
$q_0,q'_0,q_1,q'_1,\dots$ such that $q_i\in Q_i, q'_i\in Q'_i$ and
compatible to the sequence of $c_i$'s and $o_i$'s. That is to say one
has to choose $q_0\in Q_0$, a sequence of elements $u_i\in Act\backslash Act_o$
such that $q_i\step{c_i\cup o_i\cup u_i} q'_i$ and 
elements $(q_i',P_i,q_i)\in\rr$. 
The sequence $q_0,q'_0,P_0,q_1,q'_1,P_1,\dots$ summarizes a run in $\ar$
and we can determine if it is accepting, in which case Env wins the play.
Note that for the acceptance condition of $\ar$ it is relevant to know whether 
some $q\in Q$ appears in infinitely many $P_i$'s.
Therefore the set of winning plays of Env can be defined
by a {\em non deterministic} Muller automaton searching a concrete path, as we will
see below, after we make some comments.

The advantage that Env plays ``abstractly'' the game, and one selects a concrete 
path only afterward is not unfair. Again we want a controller that is secure, 
and we worry if the environment {\em could have} won. And in the case that the
controller does not have a winning strategy, it does not necessarily mean that
the environment has one, but it means that there is a risk that the environment
wins. This is related to the fact that games of incomplete information are not
determined in general: it is possible that no player has a winning strategy.

We now describe the automaton defining the set of winning plays 
and then the arena in more details. Note that the sequence 
$Q_0,Q'_0,Q_1,Q'_1,\dots$ above is uniquely determined by the sequence 
$o_0,c_0,o_1,c_1,\dots$ of actions chosen
by Cont and Env. The state space of the automaton $\ar_{ Win}$ 
recognizing the winning plays for Env is $Q\ti 2^Q$.
For all $P\neq\vide$
there is a transition $(q,P) \step{c\cup o} (q',\vide)$ if and only if 
$\exists u\inc Act\backslash Act_o,\
	\exists\, q \step{c\cup o\cup u} q'\mbox{ in }\ar$ 
and 
there is a transition $(q',\vide) \step{\epsilon} (q,P)$ if and only if 
$\exists\, (q',P,q)\in\rr$.

The automaton $\ar_{ Win}$  
non-deterministically guesses a run in $\ar$ conforming to the sequence 
$o_0,c_0,o_1,c_1,\dots$ The acceptance condition of $\ar_{ Win}$ is the
same as those of $\ar$: it can be seen as a Muller condition depending on the 
states appearing infinitely often in a run. It is given by a set of sets
$\fr\inc 2^Q$. The usual way to handle such a 
non-deterministic Muller automaton is to transform it into a non-deterministic
B{\"u}chi automaton \cite[Ch. 1]{Graedel&Th&W02}. The B{\"u}chi automaton $\br_{ Win}$ simulates
$\ar_{ Win}$ and 
guesses at some point which subset of states are going to be visited
infinitely often and that other states are no longer visited. 
The state space of $\br_{ Win}$ is $Q\cup Q\ti \fr\ti (Q\cup\{q_f\})$. It
checks in turn that each state of the chosen acceptance component $F\in\fr$ is
visited infinitely often and it is not necessary to remember the whole 
$(q,P)\in Q\ti 2^Q$ of $\ar_{ Win}$.
Using e.g. Safra's construction \cite[Ch. 3]{Graedel&Th&W02} one can transform
the B{\"u}chi automaton $\br_{ Win}$ into a {\em deterministic} Rabin automaton 
$\ccc_{ Win}$. Then the Index
Appearance Record  allows to have a deterministic
parity automaton $\dr_{ Win}$ \cite[p.86]{Graedel&Th&W02} \cite{Loeding98}. 

For defining the arena, we see that Cont and Env essentially choose the actions
$c_i$ and $o_i$:
\debeqno
	V_{Env}  =  2^{Act_c}, \quad V_{Cont} =  2^{Act_o \cap Act_{nc}}, \quad
	G	 =  (V_{Env}\ti V_{Cont}) \cup (V_{Cont}\ti  V_{Env})
\fineqno
Now the product of the arena $(V_{Env},V_{Cont},G)$ by the parity automaton 
$\dr_{ Win}$ gives rise to a parity game on a finite graph. One can determine the
winner and compute a positional winning strategy \cite[Ch.6,7]{Graedel&Th&W02}
\cite{jurdz06}. Due to the 
synchronization between the arena and $\dr_{ Win}$, the set $V_{Env}$ can
be merged to a single vertex: it is not needed to remember the move of Cont
because its effect on $\dr_{ Win}$ is sufficient. In fact the successive
sets $Q_0,Q'_0,Q_1,Q'_1,\dots$ of the above description are computed by 
$\dr_{ Win}$ (thanks to Safra's construction already in $\ccc_{ Win}$).

\begin{theorem}\label{thm-cont}
The control problem defined in Section~\ref{sec-CProblem} can be solved in 
\twoexptime. Moreover if a controller exists, then
there is one with finite memory of double exponential size.
\end{theorem}
The complexity is measured in the number $|Q|$ of states of 
$\ar=\ar_\sr \times \ar_{\neg \p}$. 
Recall that the usual control problem is \twoexptime-complete~\cite{Pnueli&Rosner89} in the size of the system and the length of the formula.

See Appendix for the proof.
The idea is to prove the following facts.
If the game $\gr$ is won by Cont then a 
controller for $\pair{\mathcal{S}}{\aformulabis}$ exists, and it can be 
constructed. Conversely if a controller for $\pair{\mathcal{S}}{\aformulabis}$ 
exists then $\gr$ is won by Cont.
By construction a strategy for Cont in $\gr$ is a finite state automaton
with expected properties about (un)observable and (un)controllable actions.
Moreover if that strategy is winning, it defines a controller for 
$\pair{\mathcal{S}}{\aformulabis}$: every run of 
$\mbox{\it lift}_{k}({\mathcal C}) \; \times \;{\mathcal S}$ fulfills
$\aformulabis$.
Conversely, if a controller for $\pair{\mathcal{S}}{\aformulabis}$ exists,
possibly with infinite memory, then this controller provides a winning
strategy for Cont in $\gr$. From the analysis above we know that if there 
is a controller for $\pair{\mathcal{S}}{\aformulabis}$, then there is one
with finite memory, and one can compute it.

%
%
\section{Example} \label{sec:4}
%
%
We illustrate our construction by a (slightly modified) example from \cite{DemriNowak05}.
The system is a bouncing ball with three actions 
$\mbox{\it lift-up}$, $\mbox{\it bounce}$ and $\mbox{\it stop}$,
where only $\mbox{\it lift-up}$ is controllable, and
only $\mbox{\it stop}$ and $\mbox{\it lift-up}$ are observable.
The law of the ball is described by
the following \logic{$\omega^2$} formula:
\debeqno
   \aformula =
   \always{\omega^2}(\mbox{\it lift-up} \; \Rightarrow \; \next{1} 
    (\always{\omega} \mbox{\it bounce} \; \wedge \; \next{\omega} \mbox{\it stop}))\ .
\fineqno
Informally, $\aformula$ states that when the ball is lifted-up, it
bounces an infinite number of times in a finite time and then stops.
Equivalently the behavior of the system is modeled by the following 
ordinal automaton of level $2$.
\\
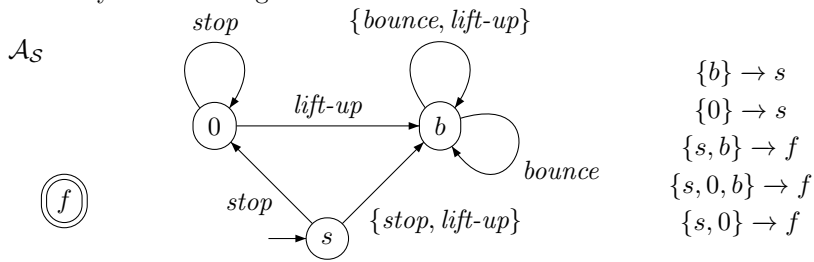
\begin{picture}(120,35)(10,-4)
\gasset{Nadjust=wh,Nadjustdist=2}

\node[Nframe=n](titre)(20,25){$\aautomaton_\sr$}
\node[Nframe=n](limites1)(115,22){$\{b\}\step{} s$}
\node[Nframe=n](limites2)(115,17){$\{0\}\step{} s$}
\node[Nframe=n](limites3)(115,12){$\{s,b\}\step{} f$}
\node[Nframe=n](limites3)(115,7){$\{s,0,b\}\step{} f$}
\node[Nframe=n](limites4)(115,2){$\{s,0\}\step{} f$}

\node[Nmarks=i,iangle=180](s)(60,0){$s$}
\node(0)(45,15){$0$}
\node(b)(75,15){$b$}
\node[Nmarks=r](f)(25,5){$f$}

\drawedge(0,b){$\mbox{\it lift-up}$}
\drawedge(s,0){$\mbox{\it stop}$}
\drawedge[ELside=r,ELpos=60](s,b){$\{\mbox{\it stop},\mbox{\it lift-up}\}$}
\drawloop[loopangle=90,ELpos=50](0){$\mbox{\it stop}$} 
\drawloop[loopangle=-20,ELpos=50](b){$\mbox{\it bounce}$}
\drawloop[loopangle=90,ELpos=50](b){$\{\mbox{\it bounce},\mbox{\it lift-up}\}$}
\end{picture}
\\
The specification is given by the \logic{$\omega^2$} formula:
\debeqno
   \aformulabis = \always{\omega^2} \next{1} \mbox{\it bounce}
\fineqno
Informally, $\aformulabis$ states that the ball should almost always be bouncing.
In the following picture of the automaton $\aautomaton_{\neg \aformulabis}$,
the star ($*$) stands for any subset of actions of $\mbox{\it Act}$.
\\
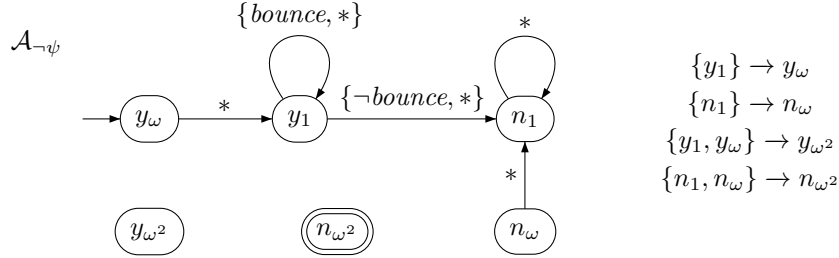
\begin{picture}(120,35)(25,-4)
\gasset{Nadjust=wh,Nadjustdist=2}

\node[Nframe=n](titre)(35,25){$\aautomaton_{\neg \aformulabis}$}
\node[Nframe=n](limites1)(130,22){$\{y_1\}\step{} y_\omega$}
\node[Nframe=n](limites2)(130,17){$\{n_1\}\step{} n_\omega$}
\node[Nframe=n](limites3)(130,12){$\{y_1,y_\omega\}\step{} y_{\omega^2}$}
\node[Nframe=n](limites4)(130,7){$\{n_1,n_\omega\}\step{} n_{\omega^2}$}

\node[Nmarks=i,iangle=180](y1)(50,15){$y_\omega$}
\node(y)(70,15){$y_1$}
\node(n)(100,15){$n_1$}
\node(n1)(100,0){$n_\omega$}
\node(y2)(50,0){$y_{\omega^2}$}
\node[Nmarks=r](n2)(75,0){$n_{\omega^2}$}

\drawedge(y1,y){$*$}
\drawedge(y,n){$\{\neg\mbox{\it bounce},*\}$}
\drawedge(n1,n){$*$}
\drawloop[loopangle=90,ELpos=50](y){$\{\mbox{\it bounce},*\}$}
\drawloop[loopangle=90,ELpos=50](n){$*$}
\end{picture}
\\
The automaton $\ar=\ar_\sr \times \ar_{\neg \p}$ is then
\\
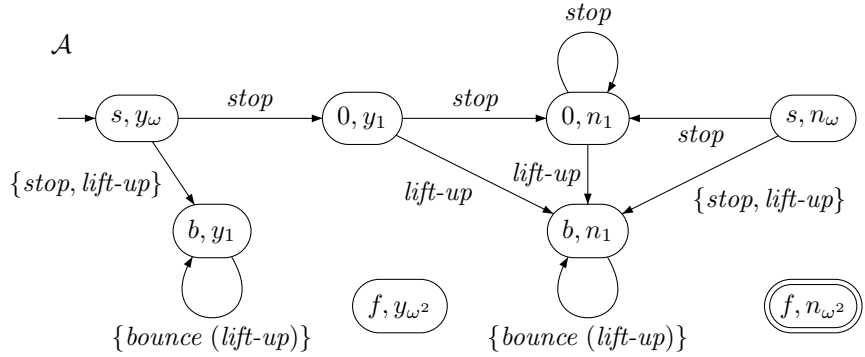
\begin{picture}(120,50)(20,-18)
\gasset{Nadjust=wh,Nadjustdist=2}

\node[Nframe=n](titre)(30,25){$\aautomaton$}

\node[Nmarks=i,iangle=180](sy1)(40,15){$s,y_\omega$}
\node(0y)(70,15){$0,y_1$}
\node(0n)(100,15){$0,n_1$}
\node(sn1)(130,15){$s,n_\omega$}
\node(by)(50,0){$b,y_1$}
\node(bn)(100,0){$b,n_1$}

\node(fy2)(75,-10){$f,y_{\omega^2}$}
\node[Nmarks=r](fn2)(130,-10){$f,n_{\omega^2}$}

\drawedge(sy1,0y){$\mbox{\it stop}$}
\drawedge[ELside=r,ELpos=20](sy1,by){$\{\mbox{\it stop},\mbox{\it lift-up}\}$}
\drawedge(0y,0n){$\mbox{\it stop}$}
\drawedge[ELside=r,ELpos=40](0y,bn){$\mbox{\it lift-up}$}
\drawedge(sn1,0n){$\mbox{\it stop}$}
\drawedge[ELpos=30](sn1,bn){$\{\mbox{\it stop},\mbox{\it lift-up}\}$}
\drawedge[ELside=r,ELpos=50](0n,bn){$\mbox{\it lift-up}$}
\drawloop[loopangle=90,ELpos=50](0n){$\mbox{\it stop}$}
\drawloop[loopangle=-90,ELpos=50](by){$\{\mbox{\it bounce}\ (\mbox{\it lift-up})\}$}
\drawloop[loopangle=-90,ELpos=50](bn){$\{\mbox{\it bounce}\ (\mbox{\it lift-up})\}$}
\end{picture}
\\
We omit here the limit transitions. In the relation $\rr\inc Q\ti 2^Q\ti Q$ the relevant
elements are 
\debeqno
(\lc b,y_1\rc,\{\lc b,y_1\rc\},\lc s,y_\omega\rc) & \qquad & 
(\lc 0,y_1\rc,\{\lc 0,n_1\rc\},\lc s,n_\omega\rc) \\
(\lc b,n_1\rc,\{\lc b,n_1\rc\},\lc s,n_\omega\rc) & &
(\lc 0,n_1\rc,\{\lc 0,n_1\rc\},\lc s,n_\omega\rc) \\
& & 
(\lc 0,n_1\rc,\{\lc 0,n_1\rc,\lc b,n_1\rc\},\lc s,n_\omega\rc)
\fineqno
If we construct the automaton $\ar_{ Win}$, we see that its (Muller) acceptance 
condition can be reduced to a B{\"u}chi condition.
In the next figure the automaton $\dr_{ Win}$ is simplified, and some unnecessary
transitions are omitted.
\\
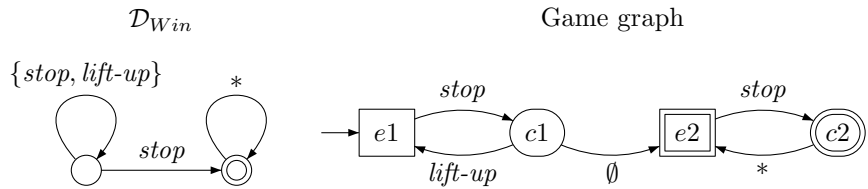
\begin{picture}(120,30)(25,10)
\gasset{Nadjust=wh,Nadjustdist=2}

\node[Nframe=n](titre)(50,35){$\dr_{ Win}$}
\node[Nframe=n](titre)(110,35){Game graph}

\node(0)(40,15){}
\node[Nmarks=r](1)(60,15){} 

\drawedge(0,1){$\mbox{\it stop}$}
\drawloop[loopangle=90,ELpos=50](0){$\{\mbox{\it stop},\mbox{\it lift-up}\}$}
\drawloop[loopangle=90,ELpos=50](1){$*$}

\node[Nmr=0,Nmarks=i](e1)(80,20){$e1$}
\node(c1)(100,20){$c1$}
\node[Nmr=0,Nmarks=r](e2)(120,20){$e2$}
\node[Nmarks=r](c2)(140,20){$c2$}

\drawedge[curvedepth=3](e1,c1){$\mbox{\it stop}$}
\drawedge[curvedepth=3](c1,e1){$\mbox{\it lift-up}$}
\drawedge[curvedepth=-3,ELside=r](c1,e2){$\vide$}
\drawedge[curvedepth=3](e2,c2){$\mbox{\it stop}$}
\drawedge[curvedepth=3](c2,e2){$*$}
\end{picture}
\\
The winning strategy for Cont is: from $c1$ always go to $e1$. The corresponding 
controller for $\pair{\mathcal{S}}{\aformulabis}$ has essentially
two loops on its initial state:
one labeled $\{\mbox{\it stop},\mbox{\it lift-up}\}$ and one labeled 
$\{\mbox{\it lift-up}\}$.

%
%
\section{Perspectives}
%
%

It is open whether the upper bounds of Theorem~\ref{thm-cont} are tight,
and whether one can find LTL-fragments or restrictions on the physical 
system such that the complexity of the control problem is lower.

We would like to extend the previous results in two directions: to timed
systems and to other linear orderings.
Given a timed automaton, it is possible to determine whether it has
Zeno behaviors. Our motivation is to extend the semantics such that after
$\w$ transitions there is a limit transition to a new control state and the
new clock values are the limit of the former ones (see \cite{BerardPicaronny}).

A Zeno behavior is not necessarily an ordinal sequence, it can be a more
general linear ordering (see \cite{BesCarton05}). One should extend the 
results to this more general class of automata.

\supr{
Automates temporis{\'e}s : produit avec l'automate qui assure un temps non nul avant
la prochaine transition. Noter qu'avec les automates temporis{\'e}s on peut aussi 
avoir plein d'autre Zeno : en temps nul, vers la gauche. Voir mes notes agraff{\'e}es
}

\paragraph{Acknowledgments.}
Great thanks to St{\'e}phane Demri and David Nowak for many interesting discussions,
helpful comments on previous versions and for their help.

\newcommand{\etalchar}[1]{$^{#1}$}

%
%
\newpage  
\section*{Appendix}
%
%

\paragraph{Correctness.}
We claim that the game $\gr$ is won by Cont iff a controller for 
$\pair{\mathcal{S}}{\aformulabis}$ exists.

If $\gr$ is won by Cont, we can compute a positional winning strategy for Cont.
It consists for each position of Cont to have exactly one outgoing edge. Now one 
can remove these intermediate states and get a finite automaton (of size 
$|\dr_{ Win}|$) where the transitions are labeled by letter in $2^{Act_{o}}$.
This automaton is a controller $\mathcal{C}$ for $\pair{\mathcal{S}}{\aformulabis}$. 
It fulfills condition (obs) of Section~\ref{sec-CProblem} clearly by construction, 
and condition (unc) because Cont chooses only controllable actions. Moreover the 
language accepted by $\mathcal{C}$ is disjoint from those of $\dr_{ Win}$ and thus
from those of $\ccc_{ Win}$, $\br_{ Win}$ and $\ar_{ Win}$. Finally the language of
$ \mbox{\it lift}_{k}({\mathcal C}) \; \times \;
{\mathcal S} \; \times \; \aautomaton_{\neg \aformulabis}$ is empty.

Conversely suppose that there exists a controller $\mathcal{C}$ for 
$\pair{\mathcal{S}}{\aformulabis}$, possibly with infinite memory. The
emptiness of $\mbox{\it lift}_{k}({\mathcal C}) \; \times \; \ar$ is
equivalent to The emptiness of $\mathcal{C}\times \ar_{ Win}$ and of
$\mathcal{C}\times \dr_{ Win}$. It follows that $\mathcal{C}$ defines a 
winning strategy in the game $\gr$.

\paragraph{Complexity.}
The sizes, in number of states, are as follows:
\debeqno
|\ar_{ Win}| & = & |Q| \\
|\br_{ Win}| & = & \bigO\left(|Q|^2.|\fr|\right) = \bigO\left(|Q|^2.2^{|Q|}\right) \\
|\ccc_{ Win}| & = & 2^{\bigO \left( |\br_{ Win}|.\log(|\br_{ Win}|) \right)} =
2^{\bigO \left( |Q|^3.2^{|Q|} \right) }
\fineqno
But the number of Rabin pairs of the acceptance condition of $\ccc_{ Win}$ is
in  $\bigO\left(|\br_{ Win}|\right)$. 
\debeqno
|\dr_{ Win}|  & = & |\ccc_{ Win}|. 2^{\bigO(|\br_{ Win}|.\log(|\br_{ Win}|))}\ \ \ 
\mbox{ hence }\ \ \ 
|\dr_{ Win}|  = 2^{\bigO \left( |Q|^6.4^{|Q|} \right) }
\fineqno
The size of $\dr_{ Win}$ is exponential only in the number of Rabin pairs of 
the acceptance condition of $\ccc_{ Win}$.
The number of priorities of the parity automaton $\dr_{ Win}$ is in
$\bigO(|\br_{ Win}|)$.
Now the number of vertices of the game graph is 
\debeqno
   n=|\dr_{ Win}|.(|V_{Cont}|+1) =  2^{\bigO\left( |Q|^3.2^{|Q|}\right) }.2^{Act_o \cap Act_{nc}}
\fineqno
the number of edges is 
\debeqno
   m=|\dr_{ Win}|.|V_{Cont}|.(|V_{Env}|+1)
\fineqno
and the number of priorities 
\debeqno
  d=\bigO(|\br_{ Win}|)\ .
\fineqno
The number of priorities of the parity game is very low compared to the number
of states. In such a case the best known deterministic algorithm for solving parity 
games is polynomial 
in the size of the graph, and exponential in the number of priorities, see 
\cite{jurdz06} and references therein. The time complexity is in:
\debeqno
   \bigO \left(d.m.\left( \frac{2n}{d} \right)^{d/2}\right)
\fineqno
which is here in 
\debeqno
   |V_{Env}|\left( 2^{\bigO\left(|Q|^6.4^{|Q|}\right)} |V_{Cont}|\right)
^{\bigO\left( |Q|^2.2^{|Q|} \right)}\ =\ \\
   |V_{Env}| 2^{\bigO\left(|Q|^8.8^{|Q|}\right)} |V_{Cont}|^{\bigO\left( |Q|^2.2^{|Q|} \right)}\ =\ \\
2^{|Act_c|} 2^{\bigO\left(|Q|^8.8^{|Q|}\right)} 2^{\bigO\left({|Act_o\cap Act_{nc}|}.|Q|^2.2^{|Q|} \right)}
\fineqno
The result of the algorithm is a positional winning strategy for the winner. In other words
it is a finite graph also with $n$ vertices. In the case that Cont wins the game, it
defines directly a controller for $\pair{\mathcal{S}}{\aformulabis}$ with at most $n$
states. More precisely the transitions of the controller are labeled by letters from
$2^{Act_o}$ and we do not need the intermediate states representing the moves of Env,
so the controller has at most $|\dr_{ Win}|$ states and 
$|\dr_{ Win}|.2^{|Act_o\cap Act_{nc}|}$ transitions.

\end{document}